# I3T: Intensity Interferometry Imaging Telescope

Pierre-Marie Gori,[1]⋆ Farrokh Vakili,[1,4]† Jean-Pierre Rivet,[1] William Guerin,[2] Mathilde Hugbart,[2]
Andrea Chiavassa,[1] Adrien Vakili,[3] Robin Kaiser,[2] Guillaume Labeyrie[2]

[1]*Université Côte d'Azur, Observatoire de la Côte d'Azur, CNRS, Laboratoire Lagrange, France*
[2]*Université Côte d'Azur, CNRS, Institut de Physique de Nice, France*
[3]*Department of Computer Science, University of Copenhagen, Denmark*
[4]*Department of Physics, Shahid Beheshti University, G.C., Tehran, Iran*

## ABSTRACT

We propose a new approach, based on the Hanbury Brown and Twiss intensity interferometry, to transform a Cherenkov telescope to its equivalent optical telescope. We show that, based on the use of photonics components borrowed from quantum-optical applications, we can recover spatial details of the observed source down to the diffraction limit of the Cherenkov telescope, set by its diameter at the mean wavelength of observation. For this, we propose to apply aperture synthesis techniques from pairwise and triple correlation of sub-pupil intensities, in order to reconstruct the image of a celestial source from its Fourier moduli and phase information, despite atmospheric turbulence. We examine the sensitivity of the method, i.e. limiting magnitude, and its implementation on existing or future high energy arrays of Cherenkov telescopes. We show that despite its poor optical quality compared to extremely large optical telescopes under construction, a Cherenkov telescope can provide diffraction limited imaging of celestial sources, in particular at the visible, down to violet wavelengths.

## 1 INTRODUCTION

50 years ago Antoine Labeyrie introduced the technique of speckle interferometry (Labeyrie 1970) to attain the diffraction limit of ground-based optical telescopes, overcoming the degrading effects of atmospheric turbulence, which limits the image sharpness of astronomical sources to atmospheric seeing limit, *i.e.* 1 arc-second typically. Along with his collaborators he published a 1st list of stars they observed with the Palomar 5 m telescope (Gezari et al. 1972). Soon after that Hanbury Brown and his team delivered a catalog of 32 fundamental stellar parameters obtained with the Narrabri intensity interferometer (Hanbury Brown et al. 1974). Intensity Interferometry is based on the Hanbury Brown and Twiss effect (HBT hereafter) (Hanbury Brown & Twiss 1956) that relates the bunching of photons in the intensity correlation between two separate telescopes to the spatial Fourier transform of the star intensity distribution projected across the sky. The advent of modern amplitude interferometry (Labeyrie 1975), with its superior sensitivity over HBT obliterated efforts to construct a 2nd generation intensity interferometer, as originally planned by Hanbury Brown (Hanbury Brown 1974). Today, optical interferometers attain hectometric baselines that reveal complex structures of stellar sources from milli-arcsecond resolutions and aperture synthesis techniques (Mourard & et al. 2015; Baines et al. 2018).

On the other hand, very large Cherenkov telescope arrays, primarily built for high-energy astrophysics (Deil et al. 2008; Mazin & et al. 2014) have motivated several international teams to reconsider

HBTII (HBT Intensity Interferometry) in the context of the future CTA (Cherenkov Telescope Array) (Dravins et al. 2013). The future full-fledged CTA will offer several thousands of baselines up to 2.5 km and a total collecting surface of about 10000 m square. Thanks to its greatly reduced sensitivity to atmospheric degrading effects, fast single photon detectors, multi-spectral channels correlation and fast digital correlators, HBTII will be a serious challenger to amplitude interferometers, specially when sub-milli-arcsecond (mas) resolutions down to 10 micro-arcsecond (*µas*) are needed at visible wavelengths. Such resolutions can potentially open a new era of discovery, e.g. measuring the angular diameter of the brightest white dwarfs (Trippe et al. 2014), of massive early-type stars in the Magellanic Clouds, and the bright line regions (BLR) components of the closest quasars (Eracleous & Halpern 1994).

Following our successful detection of temporal and spatial bunching of stellar photons on bright sources with a 1 m telescope (Guerin et al. 2017) and with two 1 m telescopes (Guerin et al. 2018), we recently determined the geometry and distance of the luminous blue variable (LBV) super-giant P Cygni in the light of its Hα emission line (Rivet et al. 2020) using a 1nm narrow band filter. HBTII has also been successfully implemented on Imaging Atmospheric Cherenkov Telescopes (IACT) specially very recently on the MAGIC (Acciari et al. 2020) and VERITAS arrays (Abeysekara & et al. 2020).

The rather fast progress on the renewal of HBTII has pushed our team to consider alternative techniques and concepts for the operation of interferometric Cherenkov arrays, starting with individual Cherenkov telescopes themselves regarded as intensity interferometers. The present paper demonstrates that, assuming a number of modifications, it is possible to transform a segmented single Cherenkov photon bucket into the equivalent of an optical telescope with an angular resolution $\sim \lambda/D$ limited by diffraction, where $\lambda$ is the observing wavelength and $D$ the telescope diameter.

⋆ E-mail: E-mail: pierre-marie.gori@oca.eu
† E-mail: E-mail: farrokh.vakili@oca.eu

In the followings we first describe the conceptual design of such an instrument: its opto-mechanical set-up, the detectors and the digital correlator. We will develop the latter point as it has crucial importance for on- and off-line processing of data and its quality. Finally, to be meaningful in terms of astrophysical outcome, we examine the signal-to-noise ratio (SNR) aspects and the limiting magnitude. Based on this aspect we also consider image reconstruction with our proposed concept down to the diffraction limit, from modulus and closure phase quantities. We call our concept Intensity Interferometric Imaging Telescope (I3T), inspired to us by Antoine Labeyrie's comparison of stellar speckle and HBT intensity interferometries (Labeyrie 1970). I3T could be implemented on the largest operating or foreseen Cherenkov telescopes such as HESS2II and the Large-Sized Telescopes (LST) of the CTA array. Finally we discuss a few technical aspects of our innovative concept and conclude on how it could contribute to stellar and extra-galactic astrophysics.

## 2 CONVERTING A PHOTON BUCKET TO A DIFFRACTION-LIMITED OPTICAL TELESCOPE

### 2.1 Principle of I3T

Extremely Large Cherenkov Telescopes (ELCTs hereafter, by reference to extremely large telescopes, ELTs), such as the twin 17 m collectors of MAGIC (Mazin et al. 2014), the HESSII telescope with its $24 \times 32$ m$^2$ primary mirror (Deil et al. 2008), the CTA 23 m LSTs (Cortina 2019) and MACE in India (Singh & Yadav 2020) made of several hundreds of elementary mirrors, also called facets, are comparable in size and collecting area to the optical ELTs under construction (Tamai et al. 2020). They benefit from fast pointing and tracking properties, but their optical quality is very poor, since their point-spread function (PSF) contains 80% of encircled energy in 1 mrad. Compared to imaging optical telescopes, that are limited to the seeing without adaptive optics, an ELCT is a simple photon bucket. The Cherenkov LST PSF is optimized to record $\gamma$ ray showers that spread over 0.2 to 1 degree across the IACT's field of view. It also results from pointing the optical axis of each facet towards a common focal plane that their focal spots pile up as a single PSF, comparable to a single facet spread function at best. In comparison, on-axis 1.5 m and 23 m parabolic mirrors that respect the Rayleigh criterion would present typically 0.33 $\mu$rad and 0.02 $\mu$rad PSFs respectively at 500 nm. However, both of them would be limited in practice to the atmospheric seeing of 1", i.e. 4.8 $\mu$rad.

Transforming an ELCT to an imaging device by means of HBTII needs additional modifications and set-ups. First, the folded PSFs from the primary *facets* need to be unfolded in order to record the photon fluxes from each facet independently. This allows for pairwise and triplewise correlations computation and can be done by tip-tilting each facet, *i.e.* redirecting its optical axis towards a given detector across the primary focal plane array of detectors (Fig. 1). This should be straightforward, since Cherenkov telescopes are equipped with hundreds and even thousands of detectors at their primary focal plane in order to collect showers from off-axis sources with distant impact points over a large field of view, typically several degrees across the sky. However, a necessary condition is that the tip-tilt drivers mounted on the mechanical support of the primary mirror dish must have enough stroke to deviate the optical axis of a given facet in the direction of a dedicated detector. At present, each such detector is 18 mm in diameter, but 2 neighbour detectors are 50 mm distant side to side in a hexagonal shape funnel, as specified for CTA's LST. Pair or triple correlation of intensities between facet mirrors provide respectively the Fourier power spectrum and the bi-spectrum of the source intensity across the sky. Thus an image of the observed source can be reconstructed from moduli and closure phase quantities of its Fourier transform (Malvimat et al. 2014).

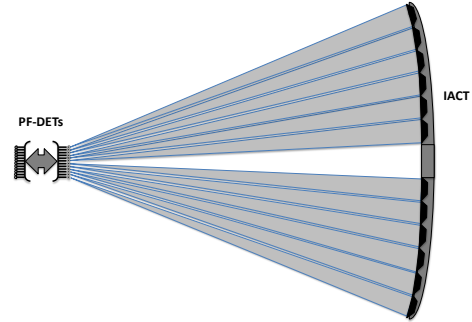

**Figure 1.** Layout of the optical and signal focal combiner of I3T. The spherical facet mirrors, mounted on the primary dish of a segmented Cherenkov telescope IACT, are tip-tilted so that their corresponding beams each impact a single, different detector at the primary focus of the telescope. The primary focus (PF-DETs) is equipped with as many detectors as the number of segments (facets). The individual detected photons are synchronously time-tagged and recorded in an archiving cluster of hard disks for off-line pairwise and triple-wise correlations.

### 2.2 Principle of Intensity Interferometry

In order to evaluate imaging performances of I3T we recall the principles of intensity interferometry, which measures the correlation function $g^{(2)}(r, \tau)$ of the fluctuating light intensity $I(t, r)$ in the focal plane of two telescopes given by:

$$g^{(2)}(\tau, r) = \frac{\langle I(t, 0) I(t + \tau, r) \rangle}{\langle I(t, 0) \rangle \langle I(t, r) \rangle}, \tag{1}$$

where the brackets denote the average over time $t$ and $r$ is the baseline vector projected on the sky referenced to one of the two telescopes taken at 0 position. For a celestial source, this function is maximum at zero delay and separation. It decreases from 2 to 1 with the increasing time delay $\tau$ depending on the temporal coherence time $\tau_c$, which is inversely proportional to the spectral bandwidth. Its dependence on the separation $r$ between the telescopes is related to the spatial coherence of the source, i.e. the usual visibility also measured by direct stellar interferometry (Labeyrie et al. 2014), which is the normalized Fourier transform of the source brightness across the sky.

Similarly, he real term of the closure phase (Malvimat et al. 2014) can be obtained from the triple correlation $g^{(3)}(\tau, \sigma, r, s)$ (Trippe et al. 2014), defined as:

$$g^{(3)}(\tau, \sigma, r, s) = \frac{\langle I(t, 0) I(t + \tau, r) I(t + \tau + \sigma, s) \rangle}{\langle I(t, 0) \rangle \langle I(t, r) \rangle \langle I(t, s) \rangle}, \tag{2}$$

where $r$ and $s$ are the baselines projected on the sky with two telescopes positions referenced against the third telescope and $\tau$, $\sigma$ time correlation variables.

The expressions of the signal-to-noise ratios (SNR) for the two- and three-point intensity correlations are (Hanbury Brown 1974):

$$\text{SNR}g^{(2)} = \alpha n_{\text{ph}} A |\gamma_{ij}|^2 \sqrt{\frac{T \Delta f}{2}}, \tag{3}$$

and (Nuñez & Domiciano de Souza 2015):

$$\text{SNR}g^{(3)} = (\alpha n_{\text{ph}} A)^{3/2} |\gamma_{ij} \gamma_{jk} \gamma_{ki}| \Delta f \sqrt{\frac{T}{\Delta \nu}}, \tag{4}$$

where $\alpha$ denotes the transmission of the Cherenkov telescope including optics and detector efficiency, $n_{\text{ph}}$ is the number of incoming photons per unit time, per unit surface and per unit optical frequency, $A$ is the geometrical mean of the telescope surfaces if the array is composed of elements with different diameters, $T$ is the total observing time, $\Delta f$ is the electronic bandwidth of the detection chain and $\Delta \nu$ is the optical bandwidth. $\gamma_{ij}$, $\gamma_{jk}$ and $\gamma_{ki}$ are the target's visibilities for baselines between the couples of $i$, $j$ and $k$ telescopes and are estimated from $g^{(2)}$ and $g^{(3)}$ correlations.

These formula are valid under the following assumptions. First, the photon flux of the science target has to be much larger than the flux of spurious collected photons (e.g. from the sky background) or the dark count rate of the detector, but should not saturate the detectors. These conditions together set a working magnitude range. Another condition is that spurious correlations, for instance due to electronic noises, are negligible compared to the bunching to be measured. This condition favors the use of a narrow spectral filtering (even if $\Delta \nu$ doesn't appear in Eq. 3) in order to enhance the coherence time and thus the measured correlation (Guerin et al. 2018).

## 3 APPLICATION TO A 20-30 METER CLASS IACT

In order to examine the performances of the proposed I3T concept, we consider the typical case of the Large-Sized Telescope (LST) of CTA (Cortina 2019), which has roughly a circular shape. Implemented on a 23 m LST, I3T would compare, for its potential angular resolution, to a 23 m optical telescope such as the Giant Magellan Telescope (Martin 2019). Note that the first of the four LSTs is under commissioning with up to expected performances demonstrated on the Crab Nebula (CTA Observatory 2020).

### 3.1 LST primary facet configuration and baseline redundancy

The LST segmented primary mirror is composed of 198 hexagonal facets providing an overall collecting surface of 368 m² (Cortina 2019). Each hexagonal facet corresponds to a spherical mirror with 1.51 m long flat to flat hexagon. The spatial hexagonal paving is depicted in Figure 2-(left), where the central facet is missing due to shadowing by the primary focus detector assembly and operational issues. Around this hole a set of 6 facets forms the first ring, followed outwards with 12 facets and so on until a 23 m diameter ring. Some very external facets are absent, so that the paving doesn't fulfill central symmetry, rather a mirror-symmetry orthogonal to the horizontal axis of rotation of LST's mount.

Since the assemblage of the facets on the mechanical support of the primary mirror is hexagonal itself, there will be an important redundancy per pair and triplet combination of facets, from the closest pairs or triplets to the most distant ones. This allows adding up independent correlation measurements corresponding to the same baseline, increasing the corresponding SNR as the square root of the number of measurements. The number of redundancies will decrease with increasing distance between these pairs or triplets so that the SNR for measured visibilities will also decrease for increasing baselines.

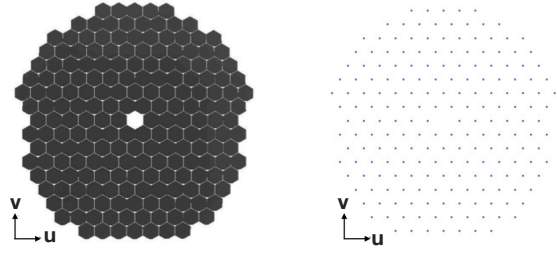

**Figure 2.** Left: geometry of the facets for the Large Size Telescope (LST, left) according to the design of CTA. Right: its digitised representation where each facet mirror is represented by a dirac distribution which has an amplitude proportional to the collecting area of a single facet.

There is no straightforward analytical way to determine the number of these redundancies, particularly when the facets assemblage deviates from a symmetrical distribution (Hayashida et al. 2015) around the optical axis (Fig. 2). A simple way to evaluate the redundancy is to use the modulation transfer function (MTF) of the array of 198 facets represented as the auto-correlation of a discrete set of identical *Dirac* distributions. Therefore the *digitized* pupil function will result in a *digitized* MTF too, with a number of Dirac's superposed due to repeated vector baselines. The hexagonal arrangement of the 198 LST facet mirrors provides 382 independent visibility samples, among the 19503 total pairwise combinations, that has been directly computed from the MTF of the digitized LST pupil.

Let us write the digitized pupil function as follows:

$$\text{P}(u,v) = \Sigma_i \delta(u - u_i, v - v_i), \tag{5}$$

where $u_i$ and $v_i$ are the coordinates of the $i$-th facet with the summation carried over the 198 facets. Now, the formal MTF is the 2D auto-correlation of $P(u,v)$, which can be rewritten as an ensemble of Diracs, each one corresponding to one of the 382 independent baselines, so that each Dirac of the MTF will be multiplied by a weighting coefficient $\rho_k$ that corresponds to the number of redundancies of the corresponding baseline:

$$\text{MTF}(u,v) = \Sigma_k \rho_k \delta(u - u_k, v - v_k), \tag{6}$$

where $u_k$ and $v_k$ are the coordinates of the spatial frequency corresponding to the baseline in the $u$-$v$ plane and the summation is carried over the 382 independent spatial frequencies. Fig. 3-top depicts the digitized MTF. We can note that the continuous hexagonal paving of facets on the LST primary aperture results in a large number of identical baselines, specially for the shortest neighbouring facets in the 3 directions parallel to the hexagons sides. As a consequence the shortest baselines sample the object power spectrum with the best SNR and the redundancies will improve the measurement proportionally to $\sqrt{\rho_k}$. Apart the zero frequency, these weights decrease from $\sqrt{180}$ at the lowest frequency of power transmissions (lowest angular resolutions), to 1 at highest frequency (highest angular resolution). Figure 3-bottom represents the azimuthal average profile of the digitized MTF and its continuous approximation $\rho_c$ based on an arcos function depending on the radial frequency $w_c \approx 46.0 \text{Mrad}^{-1}$ (Mourard et al. 1994):

$$\rho_c(\text{w}) = 2 \frac{\arccos(\frac{\text{w}}{23}) - \frac{\text{w}}{23} \sqrt{1 - (\frac{\text{w}}{23})^2}}{\pi}, \tag{7}$$

with $w = \sqrt{u^2 + v^2}$.

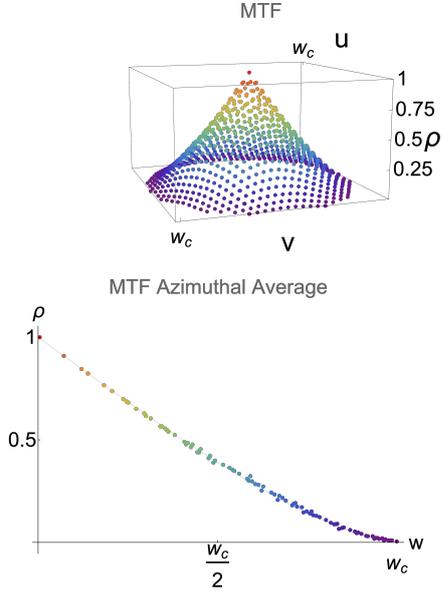

**Figure 3.** Top: the MTF of LST as the auto-correlation of its *digitized* pupil function (see Figure 2, right), with $w_c$ the LST 23 m diffraction limit cut-off frequency. Bottom: the MTF azimuthally averaged profile approximated by the following Arccos based function (see eq. 7). Both graphics are normalized to the maximum of the MTF which is 198 (number of LST facets) with the highest redundancy of 180 at the lowest spatial frequency, i.e. the shortest baseline.

## 3.2 SNR of the intensity correlation measurement and limiting magnitudes

Using equation (3) we can compute the SNR using the redundancy factor $\rho_k$ of the $k$-th baseline, for which the visibility is $\gamma_k$, and all facets are assumed having the same collecting surface $A$,

$$\text{SNRg}_k^{(2)} = \alpha n_{\text{ph}} A |\gamma_k|^2 \sqrt{\frac{\rho_k T \Delta f}{2}}. \tag{8}$$

To further simplify this equation in order to estimate the limiting magnitudes of I3T, we define:

$$\overline{\text{SNR}}\text{g}^{(2)} = \alpha n_{\text{ph}} A \sqrt{\frac{\overline{\rho} T \Delta f}{2}}, \tag{9}$$

where $\gamma$ is taken as 1 for a point-like source to estimate the limiting magnitude, $\overline{\rho} \approx 51.1$ is the value of the redundancy factor averaged over the 382 sampled spatial frequencies. In other words it is the average *power* transmission of the MTF in the $u$-$v$ plane. We consider a $\overline{\text{SNR}}\text{g}^{(2)}$ of at least 10 to obtain quality visibility values from the observations (Trippe et al. 2014) for the image reconstruction step. Fig. 4-top shows the $\overline{\text{SNR}}\text{g}^{(2)}$ dependence with the exposure time and the apparent magnitude for a point-like source. For this we also adopt the detection throughput $\alpha = 0.36$ (0.6 for the detector quantum efficiency and 0.6 for the optical throughput) and an electronic bandwidth of $\Delta f = 2$ GHz. Note that these numbers are rather optimistic compared to the presently commissioned LST, but could reasonably apply with the new generation of the SiPM-based detectors under development to upgrade the new LST camera.

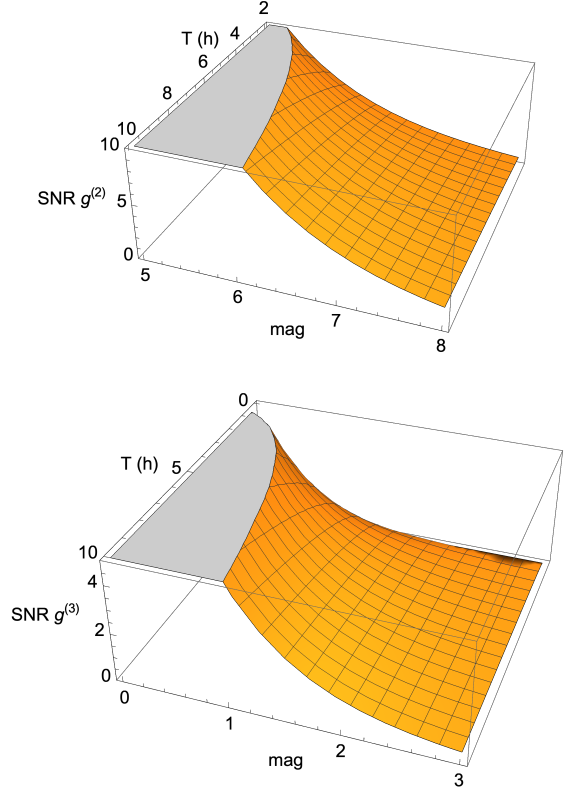

**Figure 4.** Top: SNR of an LST operated in the I3T mode for $g^{(2)}$, as a function of exposure time and magnitude. Bottom: SNR for $g^{(3)}$ as a function of the same parameters.

We can see that for an exposure time of 3h, an ELCT could achieve respectively a $\overline{\text{SNR}}\text{g}^{(2)} \approx 5$ and $\overline{\text{SNR}}\text{g}^{(2)} \approx 10$ for 6.2-mag and 5.4-mag sources. Increasing the exposure time up to 10 hours the limiting magnitude improves these values to 6.85 and 6.1. Nevertheless, at the lowest reasonable $\overline{\text{SNR}}\text{g}^{(2)}$ of 3, one may expect to observe still in 10h a 7.4 magnitude point like source. Note that for a resolved 2.5 magnitude star with a uniform disc, its second visibility maximum, i.e. 1.75%, could be measured with a $\overline{\text{SNR}}\text{g}^{(2)}$ of 5 in 10 h.

## 3.3 SNR of the closure-phase measurement

So far, we discussed different aspects to determine the visibility modulus of the Fourier transform of the source brightness projected across the sky. But if we aim to reconstruct the image of that source, we also need to compute its Fourier argument components when the source is fully resolved in order to access to its spatial details. This can be achieved by computing the triplet-wise products of the visibility, namely its bi-spectrum (Hofmann & Weigelt 1987). As already mentioned (see Sect. 2.2), only the real term of the closure phase can be accessed from triple correlation quantities. However, the imaginary part of the closure phase can be recovered for the image reconstruction step, thanks to the huge number of triple products and redundant baselines.

The accuracy of closure phase values will depend on the SNR for visibility triple products. In the same manner as for the pairwise SNR (see Eq. 3), Eq. 4 must be weighted by the bi-spectrum transfer function. Now, the numbers of pairwise and triplewise independent combinations for an interferometric array result from the classical $\frac{n(n-1)}{2}$ and $\frac{(n-1)(n-2)}{2}$ formula, so that the number of triple correlations, i.e. of closure phases, is 19306 for 198 facets of LST, very close to the 19503 baselines as already mentioned. We can thus compute the average number of redundant visibility phases as $\overline{\eta} \approx \frac{19306}{381} \approx 50.7$. 381 is the number of independent closure phases that we computed numerically from the triple correlation of pupil function, noting that closure phases do not provide the position of the source across the sky. Therefore and by analogy to Eq. 9 the average SNR of the visibility phase to estimate the limiting magnitudes for given exposure times is:

$$\overline{SNRg}^{(3)} = (\alpha n_{ph} A)^{3/2} \Delta f \sqrt{\frac{\overline{\eta} T}{\Delta \nu}}, \quad (10)$$

Fig. 4-bottom depicts the dependence of the limiting magnitude of a point-like star for given $\overline{SNRg}^{(3)}$ values as a function of the exposure time to determine the triple correlations between the facets of the LST telescope. Since both $\Delta f$ and $\Delta \nu$ appear in $\overline{SNRg}^{(3)}$, we take the same parameters as previously for $\overline{SNRg}^{(2)}$ and consider a spectral bandwidth of 0.5 nm at 780 nm which sets $\Delta \nu$ to 246.4 GHz. For a 1st magnitude target and an integration of 10h, the triple correlation can be determined with a $\overline{SNRg}^{(3)}$ whilst this $\overline{SNRg}^{(3)}$ doesn't exceed 1 for sources fainter than a 2.16 magnitude.

Let's consider the ratio of the SNRs for $g^{(3)}$ and $g^{(2)}$, which depends on the visibility and the quadratic ratio of $\Delta f$ to $\Delta \nu$, i.e. electronic over spectral bandwidths:

$$R_{SNR32} \approx \overline{|\gamma|} \sqrt{\frac{\Delta f}{2 \Delta \nu}}, \quad (11)$$

where $\overline{|\gamma|}$ is taken as the geometrical mean visibilities modulus over the 3 baselines of the $i, j$ and $k$ facets. We see that for faint objects and/or visibilities the phase closure errors will dominate the measures and in this case phaseless image reconstruction techniques such as (Fienup 1978) or more recent algorithms (Thiébaut 2008), should be preferred.

### 3.4 Imaging

The next step consists in setting the imaging capacity of I3T. For this, we apply a simple image reconstruction method to restore the intensity structures of a science target that would be fully resolved by a diffraction-limited optical telescope having the LST diameter. We take the apparent disc of a star as a relevant case, with a typical extent of 10 resolution elements (resel) across its visible hemisphere and its surface fine features. This diameter is 10 times the FWHM of an equivalent diffraction limited telescope. A red super giant (RSG), such as $\alpha$ Orionis (Betelgeuse), is a good candidate in this case since it has an angular extent of 45 mas, compared to the theoretical 4.5 mas diffraction limit of the LST in the visible. Also, Betelgeuse angular diameter varies significantly in R and I bands, from 600 to 800 nm wavelengths, in particular due to TiO absorption bands (Balega et al. 1982), so that the 45 mas value is taken here rather as a study case. The apparent I magnitude of Betelgeuse is -2.45 which is within the limits of SNR depicted in Fig. 4. Giant convective cells on the surface of Betelgeuse are inferred from spectroscopy and IR interferometry, for which intensity maps have been computed using 3D hydrodynamical simulation of stellar convection models (Chiavassa et al. 2010, 2011).

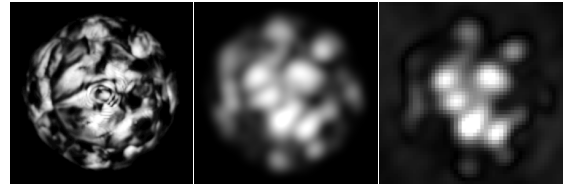

**Figure 5.** Image reconstruction results of a Red Super Giant (RSG) star and its granulation structures following the I3T concept. Left: the RSG model (580 × 580 grid) with giant convective cells on its visible disc. Middle: its theoretical image recorded by a diffraction-limited monolithic telescope with a 23 m aperture, equivalent to the LST Cherenkov telescope of CTA. Right: the RSG image recorded by an I3T after 10 hours of noiseless observations. Note that the field rotation, i.e. azimuthal coverage in the $u$-$v$ plane due to the alt-az mount of the LST telescope recovers partially the $u$-$v$ plane gaps so that the final image corresponds fairly to the diffraction-limited image (middle).

Fig. 5 depicts such a map and its image as it would be observed in the visible, say at 780 nm, by a diffraction-limited 23 m optical telescope (middle) and by a 23 m ELCT (right). The latter image is the simple inverse Fourier transform of the spatial power spectrum of the RSG, filtered by the optical transfer function of the ELCT. Here, we assume the object's Fourier power spectrum is obtained from noiseless modulus and phase data respectively, which are determined from pairwise and triple-wise intensity correlations obtained from facets focal plane signals. Indeed, the triple correlation of intensity signals give access only to the real part of closure phases (Nuñez & Domiciano de Souza 2015; Rou et al. 2013). However, combining phase closures equations where the huge number of redundancies compared to the small number of phase unknowns make possible to recover the imaginary parts of closure phases, so that a straightforward Fourier inversion of complex visibilities delivers the observed object's intensity across the sky.

In the real case however, the reconstructed image will be degraded by the finite SNR of the measurements. To mimic noisy reconstructed images we apply a simplified procedure, which only takes into account the $\overline{SNRg}^{(2)}$ measurements from (Eq. 8) and we assign as SNR for simplicity. We hamper both the real and imaginary parts of each Fourier component in the $u$-$v$ plane by corresponding noise values of this SNR. Indeed this is an optimistic scenario since the determination of phase suffers from the lower $\overline{SNRg}^{(3)}$ of Eq. 10.

In practice we first compute the noiseless Fourier spectrum of the RSG model, that we filter by the optical transfer function of the ELCT. We thus dispose of real and imaginary parts of the visibility at $u$-$v$ coordinates for $MTF(u, v) \neq 0$. Let's note $\gamma_k$ the corresponding complex visibility and $\gamma_k^{(n)}$ the noisy complex value of the visibility at the $k$-th spatial frequency. We generate the noisy power spectrum map of the object as:

$$\gamma_k^{(n)} = \gamma_k + n_k, \quad (12)$$

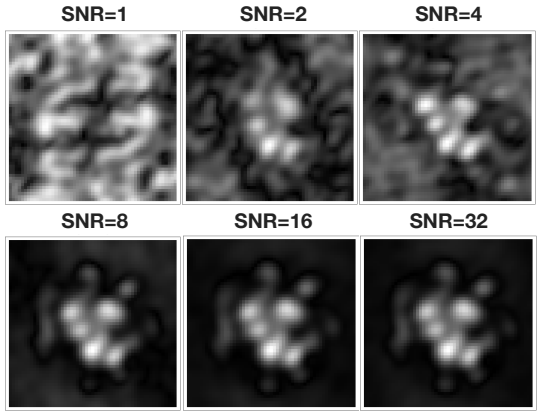

**SNR=1**    **SNR=2**    **SNR=4**

**SNR=8**    **SNR=16**    **SNR=32**

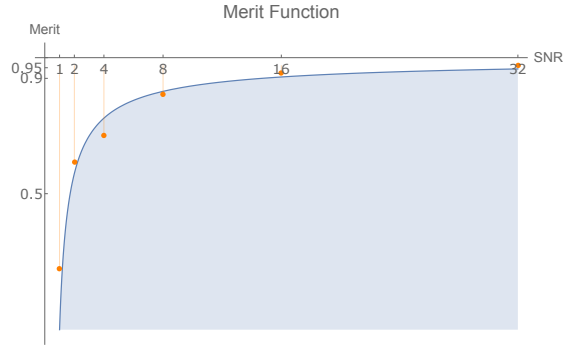

**Figure 7.** Evaluation of the I3T imaging capacity, based on the direct Fourier inversion reconstruction method: the Merit Function (MF) (Eq. 13) is plotted for increasing values (red dots) of the average SNR of visibility modulus values (Eq. 9) and computed numerically as a function of SNR (see text).

**Figure 6.** Left to right, top to bottom, reconstructed images of a RSG star photo-sphere for increasing SNR from 1 to 32. The adopted value of SNR is taken as its average over the measured spatial frequencies. It can be seen that for SNR = 1 the reconstructed image of the star has a random speckles structure, typically the size of the PSF, with a notable axisymmetric distribution. As the SNR increases, the reconstructed image of the convection giant cells at the surface of the RSG is getting closer to the RSG noiseless image recorded by a theoretically diffraction-limited optical telescope with the same diameter as the LST (see Fig. 5 middle).

where $n_k$ is a random noise that follows a normal statistical distribution according to the SNR of Eq. 8.

Fig. 6 depicts the reconstructed images of the RSG by Fourier inversion of visibilities for increasing values of SNRs. Here the SNR is defined as the average SNR over the measured spatial frequencies, as shown in Eq. (9).

The quality of the reconstructed image is already quite good for a SNR≥ 10, with in particular the clear identification of the giant convective cells at the surface of the RSG.

To be more quantitative, we compute the squared difference between the reconstructed image Image($\alpha, \delta$) and the model Model($\alpha, \delta$) as it would be recorded by the noiseless digitized pupil of the LST according to the I3T concept. The corresponding Merit Function (MF), computed as:

$$MF = 1 - \frac{\Sigma\Sigma_{\alpha,\delta}[\text{Model}(\alpha,\delta) - \text{Image}(\alpha,\delta)]^2}{\Sigma\Sigma_{\alpha,\delta}\text{Model}(\alpha,\delta)^2}, \tag{13}$$

is plotted as a function of the average SNR in Fig. 7. The continuous curve in this figure corresponds to $MF_{SNR} = 1 - \frac{3}{4}SNR^{-\frac{1}{4}}$ that represents the trend of the reconstructed image quality approximated for SNR values of 1,2,4,8,16 and 32. One sees that for SNRs larger than 10, errors on the reconstructed image is about 5% and better.

## 4 DISCUSSION

The concept of I3T (see Fig. 1) implies the tip-tilt of spherical facet mirrors on the parabolic primary mount, to have enough stroke enabling them to separately point and focus collected photons upon their dedicated detector at the common unfolded focal plane array of detectors. Now the LST has a quite open aperture, i.e. F/D=1.2 for F=28 m, which is also the focal length of individual facets, so that the linear size of their PSF is 28 mm for 80% energy concentration (Cortina & Teshima 2016). Each facet is supported on the tetrahedral primary structure, on a triangular support and can be oriented in tip-tilt by fine positioning actuators with a stroke of 36 mm and a 5 $\mu$m resolution. This stroke, given the focal distance of a facet, makes possible to deviate its PSF for about 50 cm from the on-axis focal point of LST, therefore sufficient to record separately intensity signals for each facet according to Fig. 1 and with much better sampling of the PSF with the upgraded 7420 pixels focal camera of the LST (Alispach & et al. 2020).

Besides changing the tip-tilt setpoint of the facets, the practical implementation of I3T on an LST requires an additional component to be installed in front of the focal detectors array: a specially designed mask incorporating an array of narrow band optical filters (typically 1 to 0.1 nm), and cone-shaped baffles (one per facet). The role of the optical filters is to increase the light coherence time (see Sect. 2.2). The purpose of the baffles is to reduce sky background cross-contamination. Indeed, without baffles, a given focal detector receives the target's light reflected by the facet to which it corresponds, but receives also the sky background light reflected by all facets of the dish. This spurious light is non-coherent and its residual correlation can be calibrated out, but its fluctuation would hamper severely the SNR. The detailed specification of such an intrafocal component is clearly out of the scope of this article, but its presence is mandatory to achieve the limiting magnitudes set by fundamental noise and inherent detector electronics.

To improve the I3T concept sensitivity for faint science targets, the flexibility of LST primary mirror makes possible to group a number of neighbouring facets as a single superfacet, by overlapping their focal images on the top of each other on a same detector. For instance a combination of a hexagonal set of 7 facets results in paving the 198 facets in a much smaller number of super-facets, but improves the SNR for any pair of superfacets by a factor of 7. The price to pay is a poorer sampling of the u-v plane, with much fewer spatial frequencies. Other combinations are also possible. One may consider this as a hierarchically digitized pupil of an ELCT, where facets grouping adapt to the complexity and magnitude of the observed source. Note also that the correlations can be added-up over as many nights as required to reach the necessary SNR and then carry the image reconstruction from the complex visibility estimates.

One also needs to consider the computational power needed for pair and triplet-wise intensity correlations between the 198 facets of an LST, which might sound considerably demanding. This turns out to be a moderate limitation of the I3T concept in fact. Present IACT intensity interferometers concentrate the photon flux from their entire aperture upon a single focal detector. Whereas in our case one can use time-tagged photon detector for recording the fluxes collected by different facets. Thus, we only need to store the arrival time of all photons, which is much less information: typically a few hundreds of Gbytes per facet and for 10 hours of integration time on a bright star, than what would produce a continuous signal digitized at a high rate.

Computing all the correlations between all pairs and triplets is certainly demanding but can be done off-line and thus doesn't represent a bottleneck.

Another advantage of computing the correlations off-line is that it allows to compensate for the field-rotation due to the alt-azimuthal mount of the LST. Indeed, the vector baselines between facet pairs change in orientation as projected on the sky, so that for faint sources demanding long exposures, one needs to correct for field rotation. The derotation can be done numerically, when computing the correlations, by adapting the facet pairs, corresponding to a given spatial frequency, as a function of the hour angle, before averaging the data over time.

## 5 CONCLUSION

In this paper we described a concept, called I3T, to operate a large Cherenkov telescope as an equivalent diffraction-limited optical telescope based on the technique of intensity interferometry introduced by Hanbury Brown and Twiss more than 60 years ago. The concept can be applied to short visible wavelengths, in the violet and beyond, within atmospheric transmission limits. Foreseen extremely large telescopes can hardly operate at their diffraction limit, unless laser-guide adaptive optics remove atmospheric turbulence degrading effects, or passive techniques such as classical speckle interferometry (Labeyrie 1970) are used.

We discussed a few technical and operational issues of the I3T concept for its implementation on the first LST of the CTA telescopes, recently operated on the Canarias observatory next to the MAGIC Cherenkov array. We have shown that there is no show-stopper in principle to transform an LST unit to an imaging diffraction-limited telescope. A practical implementation of our I3T concept needs however to be assessed in much more technical details, but that is beyond the scope of this paper where we focused on the principles.

Note that the full-fledged CTA array will have 4 LST units, so that if they are configured according to the I3T proposal, the signals of the four telescopes could be correlated separately and co-added to overcome the limiting magnitude of a single LST, or conversely divide by 4 the long exposure to attain the same SNR for a given magnitude.

On the other hand, if the whole CTA is operated as an intensity interferometric synthetic array, and if all the individual telescopes function in the I3T mode for their facet sub-pupils, then the u-v coverage will have much better spatial frequency resolution in the Fourier plane around a given spatial frequency defined by the baseline between 2 LSTs. Indeed facets of LST, MST and SST telescopes could be combined in pair and triplet combinations with SNRs now proportional to the geometrical mean surfaces for hybrid combinations. Hence, the u-v plane will be filled by much more resolution elements, resulting in better tracking of visibility gradients in the Fourier plane, specially around zero modulus visibilities, and more robust reconstructed image of the science object under study. Also, the I3T technique can readily be applied to existing Cherenkov telescopes like HESS28 in Namibia for instance, which presently represents the largest available collector of this type: $24 \times 30 \, m^2$ primary mosaic telescope composed of 850 flat-to-flat 90cm facets, therefore with better performances than the CTA LST collectors.

As for astrophysical programs, if I3T configuration is implemented on CTA, a number of targets will become accessible for a 23 meter imaging ELCT in the visible wavelengths. Stellar astrophysics with surface structures imaging of AGB and post-AGB stars will become practicable with reasonable total exposure times, noting that correlation quantities can be added night to night to attain the desired SNR value necessary to restore the fine details of their structures. Massive early type stars, like LBVs, Be and B[e] stars circumstellar envelopes, are favorite targets for an ELCT, as well as massive binary stars like WRs exhibiting coliding winds and emission of H, He and CNO lines signatures. Also, premain sequence stars like Ae/Be Herbig or $\beta$ Pictoris debris disks could be imaged with resolutions on the order of 4 mas, unravelled otherwise by existing or planned ELT's when we consider short wavelengths down to 400 nm.

Of special interest is the added value of I3T for extra-galactic studies, specially on globular clusters or starburst compact clusters like R136 in the LMC. The best resolved images of this object were obtained with HST and the SPHERE (Khorrami et al. 2017) on the VLT high dynamic imaging in J, H and K bands, whereas an ELCT would offer visible imaging of this object with a spatial resolution ten times better than the HST. Last but not least, the central BLR components of the closest quasars, like 3C273, could be scrutinized in Balmer emission lines such as H$\delta$, with roughly an exposure time of 10 hours according to Fig. 4. Indeed such ambitious programs will pose the serious problem of the night sky background, which is 6 to 8-mag in any 1 mrad / 4 arcmin diameter aperture due to IACTs very poor PSF optical quality. Solutions exist however to overcome this. E.g. by relaying the focal image to the Cassegrain focus of the IACT, where active optics can correct for the optical aberrations up to the required level where sky background is no more a limit for extra galactic targets. The details of such solutions are beyond the scope of the present paper that need end-to-end realistic studies.

Implementing the I3T concept on large Cherenkov telescopes, such as the South and North LSTs, HESS2, MAGIC or MACE provides a "cheaper" solution compared to optical ELT's, to obtain diffraction-limited imaging telescopes, especially at very short visible wavelengths, with reduced sensitivity to atmospheric turbulence limitations (seeing, coherence time and spectral dispersion).

Finally, it is noteworthy that, in their historical developments, homodyne interferometry came first, then its developments stopped, HBTII appeared and pulled ahead, then HBTII stopped and homodyne interferometry moved ahead again, and now HBTII may pull ahead again, at least for kilometric synthesis arrays in the visible wavelengths.

## ACKNOWLEDGEMENTS

We acknowledge the anonymous referee whose critical and constructive comments helped improving the present paper. We dedicate the I3T paper to our colleague Paul Nuñez who triggered the intensity interferometry revival on the French Riviera. This work was conducted in the context of the CTA Intensity Interferometry Working Group and reviewed by T. Mineo, J. Cortina, P. Saha, under the coordination of R. Mukherjee and T. Hassan. We thank A. Domiciano, L. Abe, O. Lai for very useful discussions; P. Berio and N. Matthews for early and critical reviews of this work before submission. F. Vakili is also indebted to W. Hoffmann for earlier exchanges on technical aspects of IACTs. This work has been supported by the CNRS Institute des Sciences de l'Univers, INSU's French CTA-section conducted by Jürgen Knödlseder from IRAP/Toulouse. We acknowledge funding from the French National Research Agency (Project I2C, ANR20-CE31-0003) and Région PACA (project I2C). PM Gori is indebted to E. Taffin de Givenchy and F. Thévenin for their support of the PhD program at UCA, Nice, France.

**DATA AVAILABILITY**

No new data were generated or analysed in support of this research.

This paper has been typeset from a TeX/LaTeX file prepared by the author.